\documentclass[12pt]{article}
\usepackage{amsmath}
\usepackage{amssymb}
\usepackage{graphicx}
\usepackage[numbers,sort&compress]{natbib}
\title{The discrete mKdV equation revisited: a Riemann-Hilbert approach
\footnotetext{}}
\author{Junyi Zhu\thanks{Email: jyzhu@zzu.edu.cn}, Xianguo Geng and Yonghui Kuang\\
{\small School of Mathematics and Statistics, Zhengzhou University,}\\
{\small Zhengzhou, Henan 450001,
People's Republic of China}}
\date{}
\begin{document}
\maketitle
\begin{abstract}
We study the plus and minus type discrete mKdV equation.
Some different symmetry conditions associated with two Lax pairs are introduced to derive the matrix Riemann-Hilbert problem with zero.
By virtue of regularization of the Riemann-Hilbert problem, we obtain the complex and real solution to
the plus type discrete mKdV equation respectively. Under the gauge transformation between the plus and minus type,
the solutions of minus type can be obtained in terms of the given plus ones.
\end{abstract}
\section{Introduction}
The discrete mKdV (dmKdV) equation \cite{AL3}
\begin{equation}\label{a1}
u_t(n,t)=\left(1\pm u^2(n,t)\right)[u(n+1,t)-u(n-1,t)],
\end{equation}
is an integrable equation in mathematical physics, and it is an important member of the discrete Ablowitz-Ladik equations \cite{AL1,AL2,AL3,AL4}.
For specific purpose, We call equation (\ref{a1}) the plus and minus type dmKdV. In this paper, we study the
plus type dmKdV equation with the help of the Riemann-Hilbert (RH) method following \cite{D-M-R},
then the solutions of the minus type can be obtained by virtue of a gauge transformation.

The plus type dmKdV equation (\ref{a1}) admits the following Lax pair formulation \cite{G-G}:
\begin{equation}\label{b1}
\psi(n+1,t)=\gamma_n(I+Q_n)Z\psi(n,t),\quad \psi_t(n,t)=(k\sigma_3+\tilde{Q}_n)\psi(n,t),
\end{equation}
where $I$ is the identity matrix, $\gamma_n=\sqrt{\det(I+Q_n)}^{-1}$, and the matrices $Q_n,Z, \tilde{Q}_n$ take the form
\begin{equation}\label{b2}
\begin{aligned}
&Q_n=\left(\begin{matrix}
0&u(n,t)\\
-u(n,t)&0
\end{matrix}\right),\quad Z=\left(\begin{matrix}
z&0\\
0&1
\end{matrix}\right),\\
&\tilde{Q}_n=Q_n+Z^{-1}Q_{n-1}Z,\quad k=\frac{1}{2}(z-z^{-1}),
\end{aligned}
\end{equation}
with $z$ is a spectral parameter. We note that the Lax pair formulation (\ref{b1}) can be rewritten as
\begin{equation}\label{a4}
\psi(n+1,t)=(I+Q_n)Z\psi(n,t),\quad \psi_t(n,t)=(k\sigma_3+\tilde{Q}_n-Q_nQ_{n-1})\psi(n,t),
\end{equation}
which are the ones in \cite{G-G}.

The plus type dmKdV equation (\ref{a1}) admits another Lax pair formulation \cite{AL3}
\begin{equation}\label{a2}
\varphi(n+1,t)=\gamma_n(E+Q_n)\varphi(n,t),\quad \varphi_t(n,t)=(\omega\sigma_3+\hat{Q}_n)\varphi(n,t),
\end{equation}
where $Q_n$ is defined as (\ref{b2}) and
\begin{equation}\label{a3}\begin{aligned}
&E=\left(\begin{matrix}
z&0\\
0&z^{-1}
\end{matrix}\right),\quad \gamma_n=\frac{1}{\sqrt{\det(E+Q_n)}},\\
&\hat{Q}_n=EQ_n+Q_{n-1}E,\quad \omega=\frac{1}{2}(z^2-z^{-2}).
\end{aligned}\end{equation}
We note that the Lax pair (\ref{a2}) can be rewritten in a similar form as (\ref{a4}) which are the ones as in \cite{AL3}.

It is known that self-dual network can also be reduced to the discrete analogue of the mKdV equation \cite{NWS}. we note that there are many other differential-difference
equations which can be transformed into the dmKdV equation \cite{NK1,PC,CC,NK2,NK3,MN,LJL,NK4}.
The dmKdV equation has widely applications in the fields as plasma physics, electromagnetic waves
in ferromagnetic, antiferromagnetic or dielectric systems, and can be solved by the method of inverse scattering transform, Hirota bilinear,
Algebro-geometric approach and others \cite{AL3,AS,BPS,LDZ,LCZ,NQC,G-G,ZD,Y-Z,WG,ZMH,WM,AAS,YZ,ZZW,NK,SZW,SC,OCM}.

In this paper, we firstly consider the Lax pair (\ref{b1}) and introduce a special symmetry condition, which imply that $u(n, t)$ in (\ref{a1})
can be extended to complex value. For simplicity, we suppose that $u(n, t)\neq\pm i$ and $\gamma_n$ is chosen as one of branches.
It is noted that the complex solution in this paper is different from the complexiton solution introduced by W.X. Ma \cite{MWX,M-M,M,M4},
in which the complexiton solutions are obtained in the sense of complex eigenvalues, and are still real.
Next we consider the linear system (\ref{a2}) under the usual symmetry condition which confine the potential $u(n, t)$ to be real.
We note that to obtain the real solution of the plus type dmKdV equation, one needs to introduce some constraint condition.
For one soliton solution as an example, we assume that $(2n+1)\eta+2\cosh2\xi\sin2\eta t=m\pi$, where $m=0,\pm1,\pm2,\cdots$
and the discrete spectrum for $N=1$ is defined as $z_1=e^{\xi+i\eta}$.

The organization of this paper is as follows. In section 2, we derive the complex solution of the plus type dmKdV equation by virtue of
RH problem associated with linear system (\ref{b1}).
In section 3, we derive the real solution of the plus type dmKdV equation by virtue of
RH problem associated with linear system (\ref{a2}). In section 4, we study the gauge transformation between the plus and minus type dmKdV equation,
from which the solution of minus type can be obtained in terms of the given plus one.

\section{Complex solution of the dmKdV equation}
\subsection{The spectral analysis}
For the sake of convenience, we write the spectral equation (\ref{b1}) in terms of the matrix
$$J(n)=\psi(n) Z^{-n}e^{-k\sigma_3t}.$$
Hence, the dmKdV equation allows the Lax representation:
\begin{equation}\label{b3}
J(n+1)=\gamma_n(I+Q_n)ZJ(n)Z^{-1},
\end{equation}
and
\begin{equation}\label{b4}
J_t(n)=k[\sigma_3,J(n)]+\tilde{Q}_nJ(n).
\end{equation}
Here and after we suppress the variables dependence for simplicity of notation.

Now we introduce matrix Jost functions $J_\pm(n,z)$ of the spectral equation (\ref{b3})
obeying the asymptotic conditions
\begin{equation}\label{b5}
J_\pm(n,z)\rightarrow I,\quad n\rightarrow\pm \infty.
\end{equation}
Then there exists the scattering matrix $S(z)$ admitting
\begin{equation}\label{b6}
J_-(n,z)=J_+(n,z)Z^nS(z)Z^{-n}, \quad S(z)=\left(\begin{matrix}
a_+(z)&-b_-(z)\\
b_+(z)&a_-(z)
\end{matrix}\right).
\end{equation}
Here we assume that the Jost functions and the scattering matrix satisfy the symmetry condition
\begin{equation}\label{b7}
J_\pm^T(n,z^-)=J_\pm^{-1}(n,z),\quad \det J_\pm(n,z)=1,
\end{equation}
and
\begin{equation}\label{b8}
S^T(n,z^-)=S^{-1}(n,z),\quad \det S(n,z)=1,\quad z^-=z^{-1}.
\end{equation}

In the following, we consider the asymptotic behavior of the solution $J(n,z)$.
To this end, we first let
\begin{equation}\label{b14}
J(n,z)=J^{(0)}(n)+z^{-1}J^{(1)}(n)+O(z^{-2}),\quad z\rightarrow\infty,
\end{equation}
and substitute it into the spectral equation (\ref{b3}). This yields
\begin{equation}\label{b15}
J_{11}^{(0)}(n+1)=\gamma_nJ_{11}^{(0)}(n),\quad J_{22}^{(0)}(n+1)=\gamma_n^{-1}J_{22}^{(0)}(n),
\end{equation}
and $J_{12}^{(0)}(n)=0$,
\begin{equation}\label{b16}
J_{12}^{(1)}(n)=-u(n)J_{22}^{(0)}(n),\quad J_{21}^{(0)}(n+1)=-u(n)J_{11}^{(0)}(n+1).
\end{equation}

Next, according to the symmetry (\ref{b7}), we let
$$J^{-1}(n,z)=\tilde{J}_{(0)}+z\tilde{J}_{(1)}+O(z^2),\quad z\rightarrow0,$$
and find similarly $\tilde{J}_{(0)21}(n)=0$,
\begin{equation}\label{b17}
\begin{aligned}
&\tilde{J}_{(0)11}(n+1)=\gamma_n\tilde{J}_{(0)11}(n),\quad \tilde{J}_{(0)22}(n+1)=\gamma_n^{-1}\tilde{J}_{(0)22}(n),\\
&\tilde{J}_{(0)12}(n+1)=-u(n)\tilde{J}_{(0)11}(n+1),\quad \tilde{J}_{(1)21}(n)=-u(n)\tilde{J}_{(0)22}(n).
\end{aligned}
\end{equation}

We will now discuss the analytic of the Jost solutions. The spectral equation (\ref{b3}), as a iterative relation, can
be written as
\begin{equation}\label{b18}
J(n,z)=\nu_+(n)\lim\limits_{N\rightarrow\infty}\prod\limits_{l=n}^N(Z^{-1}(I-Q_l))J(N+1,z)Z^{N-n+1},
\end{equation}
here and after we introduce two new functions $\nu_\pm(n)$ as following
\begin{equation}\label{b19}
\nu_+(n)=\prod\limits_{l=n}^\infty\gamma_l,\quad \nu_-(n)=\prod\limits_{l=-\infty}^{n-1}\gamma_l.
\end{equation}
We note that the first column of the matrix equation (\ref{b18}) involves two positive power series in $z$,
while the second column involves two negative power series in $z$. Thus the first column
$J_+^{[1]}(n,z)$ of the Jost function $J_+$ is analytical for $|z|<1$, denoted by ${\mathbb{C}}_I$, and
the second column $J_+^{[2]}(n,z)$ is analytical for $|z|>1$ or (${\mathbb{C}}_O$).
By the same way one can show that the column $J_-^{[1]}(n,z)$ is analytical for $|z|>1$ or (${\mathbb{C}}_O$).
We introduce a matrix function
$$\Phi_+(n,z)=\left(J_-^{[1]},J_+^{[2]}\right)$$
which is analytical in ${\mathbb{C}}_O$ and solves the spectral equation (\ref{b3}).

It follows from the symmetry condition (\ref{b7}) that the rows $(J_-)_{[1]}^{-1}$ and $(J_+)_{[2]}^{-1}$ are
analytical in ${\mathbb{C}}_I$. Thus the matrix function
$$\Phi_-^{-1}(n,z)=\left(\begin{array}{c}
(J_-)_{[1]}^{-1}\\
(J_+)_{[2]}^{-1}
\end{array}\right)$$
is analytical in ${\mathbb{C}}_I$ and solves the adjoint spectral problem of (\ref{b3}).

By virtue of the definition (\ref{b6}) of the scattering matrix , we find
\begin{equation}\label{b20}
\Phi_+(n,z)=J_\pm Z^nS_\pm Z^{-n},
\end{equation}
where
$$S_+=\left(\begin{matrix}
a_+&0\\
b_+&1
\end{matrix}\right),\quad S_-=\left(\begin{matrix}
1&b_-\\
0&a_+
\end{matrix}\right).$$
Hence, on use of (\ref{b7}), we obtain
\begin{equation}\label{b21}
\det\Phi_+=\det J_\pm\det S_\pm=a_+(z).
\end{equation}
Following the same procedure as the one used for $\Phi_+$, one obtains
\begin{equation}\label{b22}
\begin{aligned}
&\Phi_-^{-1}(n,z)=Z^nT_\pm Z^{-n}J_\pm^{-1},\quad \det\Phi_-^{-1}(n,z)=a_-(z),\\
&T_+=\left(\begin{matrix}
a_-&b_-\\
0&1
\end{matrix}\right),\qquad T_-=\left(\begin{matrix}
1&0\\
b_+&a_-
\end{matrix}\right).
\end{aligned}
\end{equation}

Asymptotic formulae for these sectionally analytic functions can be derived from equations (\ref{b14}) to (\ref{b17}),
\begin{equation}\label{b23}
\Phi_+(n,z)\rightarrow\Phi_+^{(0)}(n)=\left(\begin{matrix}
\nu_-(n)&0\\
-u(n-1)\nu_-(n)&\nu_+(n)
\end{matrix}\right),\quad z\rightarrow\infty,
\end{equation}
and
\begin{equation}\label{b24}
\Phi_-^{-1}(n,z)\rightarrow\Phi_{-(0)}^{-1}(n)=\left(\begin{matrix}
\nu_-(n)&-u(n-1)\nu_-(n)\\
0&\nu_+(n)
\end{matrix}\right),\quad z\rightarrow0.
\end{equation}
where $\nu_\pm(n)$ defined in (\ref{b19}). Indeed, to obtain equation (\ref{b23}), we know, from the definition of $\Phi_+$ and the asymptotic expansion (\ref{b14}), that
$$\Phi_+^{(0)}(n)=\left(\begin{matrix}
J_{-11}^{(0)}(n)&0\\
J_{-21}^{(0)}(n)&J_{+22}^{(0)}(n)
\end{matrix}\right),$$
Iterating the relations (\ref{b15}) and (\ref{b16}), we find
$$\begin{aligned}
&J_{-11}^{(0)}(n)=\gamma_{n-1}J_{-11}^{(0)}(n-1)=\cdots=\nu_-(n),\\
&J_{-21}^{(0)}(n)=-u(n-1)J_{-11}^{(0)}(n)=-u(n-1)\nu_-(n),\\
&J_{+22}^{(0)}(n)=\gamma_nJ_{+22}^{(0)}(n+1)=\cdots=\nu_+(n),
\end{aligned}$$
which give (\ref{b23}) in terms of the boundary condition (\ref{b5}). Equation (\ref{b24}) can be obtained from (\ref{b17}) in a same way.

We note that the symmetry condition about these sectionally analytic functions can be obtained from that of the Jost solutions as
\begin{equation}\label{b25}
\Phi_+^T(n,z^-)=\Phi_-^{-1}(n,z).
\end{equation}
In addition, equations (\ref{b20}) and (\ref{b21}) imply that $a_+(z)$ and $a_-(z)$ are analytical in the domain of
${\mathbb{C}}_O$ and ${\mathbb{C}}_I$ respectively. Furthermore they admit the following asymptotic behavior
\begin{equation}\label{b26}
a_+(z)\rightarrow \nu,~z\rightarrow\infty;\quad
a_-(z)\rightarrow \nu,~z\rightarrow0,
\end{equation}
where $\nu=\nu_+(n)\nu_-(n)=\prod_{l=-\infty}^\infty\gamma_l$.

It is noted that the potential $u(n)$ can be reconstructed by the analytic functions. Indeed, from the first equation of (\ref{b16}), we find
\begin{equation}\label{b27}
u=-\frac{J_{12}^{(1)}(n)}{J_{22}^{(0)}(n)}=-\lim\limits_{z\rightarrow\infty}\frac{(z\Phi_+)_{12}}{(\Phi_+)_{22}}=-\frac{\Phi_{+12}^{(1)}}{\Phi_{+22}^{(0)}},
\end{equation}
while the second equation of (\ref{b16}) is an identity.

\subsection{RH problem and its regularization}

Now we can introduce the RH problem
\begin{equation}\label{c1}\begin{aligned}
&\Phi_-^{-1}(n,z)\Phi_+(n,z)=Z^nG(z)Z^{-n},\quad |z|=1,\\
&G(z)=T_+S_+=T_-S_-=\left(\begin{matrix}
1&b_-(z)\\
b_+(z)&1
\end{matrix}\right).
\end{aligned}
\end{equation}
The normalization of the RH problem is given by (\ref{b23}) which is noncanonical.
Hence the dmKdV potential can be retrieved by virtue of the solution of RH problem.

In order to obtain the soliton solutions of the dmKdV equation, we take $G(z)=I$ and
suppose $a_+(z)$ has simple zeros at $z_j\in{\mathbb{C}_O},~j=1,\cdots,N$. From the symmetry
(\ref{b25}), we know that $\det\Phi_+(z_j)=0,~\det\Phi_-^{-1}(z^-_l)=0,~j,l=1,\cdots,N$.
In this case, problem (\ref{c1}) is called the RH one with zeros which can be solved by virtue of its regularization.

To obtain the relevant regular problem, we introduce a rational matrix function
$$
\chi_j^{-1}=I+\frac{z_j-z^-_j}{z-z_j}P_j, \quad P_j=\frac{|y_j\rangle\langle\tilde{y}_j|}{\langle \tilde{y}_j|y_j\rangle},
$$
where the eigenvector $|y_j\rangle$ solves $\Phi_+(n,z_j)|y_j\rangle=0$.
Since $\Phi_+(n,z_j)$ admits the linear system (\ref{b3}) and (\ref{b4}), then we have
$$\begin{aligned}
&\Phi_+(n,t,z_j)Z^{-1}(z_j)|y_j\rangle(n+1,t)=0,\\
&\Phi_+(n,t,z_j)(|y_j\rangle_t-k_j\sigma_3|y_j\rangle)(n,t)=0,
\end{aligned}$$
which imply that
\begin{equation}\label{c1a}
|y_j\rangle(n,t)=Z^n(z_j)e^{k_j\sigma_3t}|y_j\rangle_0,\quad k_j=(z_j-z_j^-)/2,
\end{equation}
where $|y_j\rangle_0$ is an arbitrary constant vector. In addition, one finds that $\langle \tilde{y}_j|=|y_j\rangle^T$ satisfies $\langle \tilde{y}_j|\Phi_-^{-1}(n,z_l^-)=0$.

Therefore the product $\Phi_+(z)\chi_j^{-1}(z)$ is regular at the point $z_j$ and $\chi_l(z)\Phi_-^{-1}(z)$ is regular at $z^-_l$, where
\begin{equation}\label{c2}
\chi_l=I-\frac{z_l-z^-_l}{z-z^-_l}P_l.
\end{equation}
The regularization of all the other zeros is performed similarly and eventually we obtain the following representation for the analytic solution
\begin{equation}\label{c3}
\Phi_\pm=\phi_\pm\Gamma,\quad \Gamma=\chi_N\chi_{N-1}\cdots\chi_1,
\end{equation}
where the holomorphic matrix functions $\phi_\pm$ solve the regular RH problem
\begin{equation}\label{c3a}
\phi_-^{-1}(n,z)\phi_+(n,z)=I.
\end{equation}

We note that the soliton matrix $\Gamma$ can be decomposed into simple fractions
\begin{equation}\label{c4}
\Gamma=I-\sum\limits_{j,l=1}^N\frac{1}{z-z^-_l}|y_j\rangle(D^{-1})_{jl}\langle \tilde{y}_l|,\quad D_{lj}=\frac{\langle \tilde{y}_l|y_j\rangle}{z_j-z_l^-}.
\end{equation}

In the following, we will establish the relationship between the solution of dmKdV equation and the soliton matrix. Taking into account the asymptotic formula (\ref{b23}) and the expression of $\Gamma$ (\ref{c4}), we choose
$\Phi_+^{(0)}=\phi_+$. Then, in view of (\ref{c3}), we find
\begin{equation}\label{c5}
\Gamma(n,z)=I+z^{-1}\Gamma^{(1)}(n)+O(z^{-2}),\quad z\rightarrow\infty.
\end{equation}
and $\Phi_{+12}^{(1)}(n)=\nu_-(n)\Gamma_{12}^{(1)}(n)$. In addition, the assumption $G(z)=I$ implies that $b_\pm(z)=0$ and then $a_+(z)a_-(z)=1$ in view of (\ref{b8}).
From (\ref{b26}), we know that $\nu=\nu_+(n)\nu_-(n)=1$.
Hence the potential $u(n)$ can be rewritten as
\begin{equation}\label{c6}
u(n)=-\frac{\nu_-(n)\Gamma_{12}^{(1)}(n)}{\nu_+(n)}=-\nu_-^2(n)\Gamma_{12}^{(1)}(n).
\end{equation}

Next we will establish the relationship between $\nu_-$ and $\Gamma$. Since $G(z)=I$, the RH problem (\ref{c1}) reduces to $\Phi_+=\Phi_-$,
from which we can consider the asymptotic behavior of $\Phi_+$ near $z=0$. Indeed, the asymptotic formulae (\ref{b22}) and (\ref{b23}) imply that
$$\Phi_+\rightarrow\Phi_{-(0)}(n)=\left(\begin{matrix}
\nu_+(n)&u(n-1)\nu_-(n)\\
0&\nu_-(n)
\end{matrix}\right),\quad z\rightarrow0.$$
Thus from (\ref{c3}) we obtain
\begin{equation}\label{c7}
\begin{aligned}
\Gamma(n,z)|_{z=0}&=(\Phi_+^{(0)})^{-1}(n)\Phi_+|_{z=0}\\
&=\left(\begin{matrix}
\nu_+^2(n)&u(n-1)\\
u(n-1)&\nu_-^2(n-1)
\end{matrix}\right),
\end{aligned}
\end{equation}
which implies that $\nu_-^2(n)=\Gamma_{22}(n+1,z=0)$. As a result,
the potential $u(n)$ takes the form
\begin{equation}\label{c9}
u(n)=-\Gamma_{12}^{(1)}(n)\Gamma_{22}(n+1,z=0).
\end{equation}

\subsection{Complex soliton solutions}
In this section, we will derive the soliton solutions of the dmKdV equation (\ref{a1}).
To this end, we let
$$z_j=e^{\xi_j+i\eta_j},~\xi_j>0, \quad |y_j\rangle_0=\left(\begin{array}{c}
e^{a_j+i\alpha_j}\\
1
\end{array}\right).$$
Hence the vector $|y_j\rangle,~(j=1,2,\cdots,N)$ take the form
\begin{equation}\label{d1}
|y_j\rangle=e^{\frac{1}{2}(\theta_j(n)+i\phi_j(n))}\left(\begin{array}{c}
e^{\frac{1}{2}(X_j(n,t)+i\varphi_j(n,t))}\\
e^{-\frac{1}{2}(X_j(n,t)+i\varphi_j(n,t))}
\end{array}\right),
\end{equation}
where
\begin{equation}\label{d2}
\begin{aligned}
&X_j(n,t)=n\xi_j+2\sinh\xi_j\cos\eta_jt+a_j,\\
&\varphi_j(n,t)=n\eta_j+2\cosh\xi_j\sin\eta_jt+\alpha_j,
\end{aligned}
\end{equation}
with $\theta_j(n)=X_j(n,0), \phi_j(n)=\varphi_j(n,0)$.

In particularly, for $N=1$, equation (\ref{c4}) reduces to
\begin{equation}\label{d3}
\Gamma(n,z)=I-\frac{z_1-z_1^-}{z-z_1^-}\frac{|y_1\rangle\langle\tilde{y}_1|}{\langle\tilde{y}_1|y_1\rangle},
\end{equation}
from which we have the complex form of one-soliton solution to dmKdV equation
\begin{equation}\label{d6}
u(n,t)=\frac{z_1^2-1}{2}{\rm sech}\{X_1(n+1,t)+i\varphi_1(n+1,t)\}.
\end{equation}

For $N=2$, we find
\begin{equation}\label{d7}
\begin{aligned}
\Gamma^{(1)}(n,z)=&-\frac{1}{\det D}\left\{D_{22}|y_1\rangle\langle\tilde{y}_1|-D_{21}|y_2\rangle\langle\tilde{y}_1|\right.\\
&\qquad\left.+D_{11}|y_2\rangle\langle\tilde{y}_2|-D_{12}|y_1\rangle\langle\tilde{y}_2|\right\},
\end{aligned}
\end{equation}
and
\begin{equation}\label{d8}
\begin{aligned}
\Gamma(n,0)=I+&\frac{1}{\det D}\left\{z_1\left[D_{22}|1\rangle\langle\tilde{1}|B-D_{21}|2\rangle\langle\tilde{1}|B\right]\right.\\
&\qquad\left.+z_2\left[D_{11}|2\rangle\langle\tilde{2}|B-D_{12}|1\rangle\langle\tilde{2}|B\right]\right\},
\end{aligned}
\end{equation}
where $\det D$ is obtained according to the definition of (\ref{c4}) and (\ref{d1}) as
\begin{equation}\label{d9}
\det D=\Xi\Omega_2(n),\quad \Xi=\left(\prod\limits_{j,l=1}^2(z_j-z_l^-)\right)^{-1}\frac{2e^{\theta_1+\theta_2+i(\phi_1+\phi_2)}}{z_1z_2},
\end{equation}
and
\begin{equation}\label{d10}
\begin{aligned}
\Omega_2(n)=&(z_1-z_2)^2\cosh\{\vartheta_1(n,t)+\vartheta_2(n,t)\}\\
&+(z_1z_2-1)^2\cosh\{\vartheta_1(n,t)-\vartheta_2(n,t)\}-(z_1^2-1)(z_2^2-1),\\
\end{aligned}
\end{equation}
with
$$\vartheta_j(n,t)=X_j(n,t)+i\varphi_j(n,t).$$
In addition, from (\ref{d1}), (\ref{d7}) and (\ref{d8}),
we know that
\begin{equation}\label{d12}
\Gamma_{12}^{(1)}(n)=-\frac{V_2(n+1)}{\Omega_2(n)},\quad \Gamma_{22}(n,0)=\frac{\Omega_2(n)}{\Omega_2(n+1)},
\end{equation}
where
\begin{equation}\label{d13}
\begin{aligned}
V_2(n)=&(z_2-z_1)(z_1z_2-1)\left[z_1(z_2^2-1)\cosh\{\vartheta_1(n,t)\}\right.\\
&\quad\left.-z_2(z_1^2-1)\cosh\{\vartheta_2(n,t)\}\right].\\
\end{aligned}
\end{equation}
Hence the solution of the plus type dmKdV equation for $N=2$ can be given by
\begin{equation}\label{d12}
u(n)=\frac{V_2(n+1)}{\Omega_2(n+1)}.
\end{equation}

It is noted that the solution $u(n)$ can also be derived through $\Gamma_{12}(n+1,z=0)$ by (\ref{c7}).
It is verified that the representations of solution by $\Gamma_{12}(n+1,z=0)$ are same as the ones in
(\ref{d6}) and (\ref{d12}).

\setcounter{equation}{0}
\section{Real solutions of the dmKdV equation}
\subsection{The spectral analysis}

In the section, we consider the inear system (\ref{a2}) and assume that the solution $u(n,t)$ is a real function. After the transformation
$$J(n)=\varphi(n)E^{-n}e^{-\omega\sigma_3t},$$
the dmKdV equation allows the Lax representation:
\begin{equation}\label{e1}
J(n+1)=\gamma_n(E+Q_n)J(n)E^{-1},
\end{equation}
and
\begin{equation}\label{e2}
J_t(n)=\omega[\sigma_3,J(n)]+\hat{Q}_nJ(n).
\end{equation}
We assume that the function $J(n,z)$ admits the following symmetry conditions
\begin{equation}\label{e3}
J^\dag(n,\bar{z})=J^{-1}(n,z),\quad \sigma_3J(n,-z)\sigma_3=J(n,z),
\end{equation}
where $\bar{z}=(z^*)^{-1}$ with $z^*$ denotes the complex conjugate of $z$.

The Jost functions $J_\pm(n,z)$ and the scattering matrix $S(z)$ can be introduced in the same way as in
(\ref{b5}) and (\ref{b6}). It is readily verified that the matrices $J_\pm(n,z)$ and $S(z)$ are
unimodular, and satisfy the symmetry conditions (\ref{e3}).

We note that similar considerations apply to the asymptotic behavior of the Jost functions $J_\pm(n,z)$, one find
\begin{equation}\label{e4}\begin{aligned}
&J(n,z)=J^{(0)}(n)+z^{-1}J^{(1)}(n)+O(z^{-2}),\quad z\rightarrow\infty,\\
&J(n,z)=J_{(0)}(n)+zJ_{(1)}(n)+O(z^2),\quad z\rightarrow0,
\end{aligned}\end{equation}
where the diagonal matrices $J^{(0)}(n)$ and $J_{(0)}(n)$ admit the following iterative relations
\begin{equation}\label{e5}
J^{(0)}(n+1)=\left(\begin{matrix}
\gamma_n&0\\
0&\gamma_n^{-1}
\end{matrix}\right)J^{(0)}(n),\quad
J_{(0)}(n+1)=\left(\begin{matrix}
\gamma_n^{-1}&0\\
0&\gamma_n
\end{matrix}\right)J_{(0)}(n).
\end{equation}
In addition, the solution can be constructed by
\begin{equation}\label{e6}
u(n)=-\frac{J_{12}^{(1)}(n)}{J_{22}^{(0)}(n)}.
\end{equation}

The analytical properties of the Jost functions $J_\pm(n,z)$ are the same as the ones in complex section above,
and can be used to define the same sectionally holomorphic $\Phi_+(n,z)$ and $\Phi_-^{-1}(n,z)$ as
(\ref{b20}) and (\ref{b22}) with $Z$ replaced by $E$. Furthermore, we have the symmetry condition
about $\Phi_\pm(n,z)$
\begin{equation}\label{e7}
\Phi_+^\dag(n,\bar{z})=\Phi^{-1}(n,z),
\end{equation}
and the asymptotic behavior
\begin{equation}\label{e8}
\Phi_+(n,z)\rightarrow\Phi_+^{(0)}(n)=\left(\begin{matrix}
\nu_-(n)&0\\
0&\nu_+(n)
\end{matrix}\right),\quad z\rightarrow\infty,\\
\end{equation}
and
\begin{equation}\label{e9}
\Phi_-^{-1}(n,z)\rightarrow\tilde{\Phi}_-^{(0)}(n)=\left(\begin{matrix}
\nu_-(n)&0\\
0&\nu_+(n)
\end{matrix}\right),\quad z\rightarrow0,
\end{equation}
where the real functions $\nu_\pm(n)$ are defined as in (\ref{b19}).

\subsection{The regularization of the RH problem and the soliton solutions}
The RH problem associated with $\Phi_\pm(n,z)$ can be constructed as in (\ref{c1}),
while the normalization of the RH problem is given by (\ref{e8}).

In order to obtain the real soliton solutions of the dmKdV equation, we take $G(z)=I$ and
suppose $a_+(z)=\det\Phi_+(n,z)$ has simple zeros at $\pm z_j\in{\mathbb{C}_O},~j=1,\cdots,N$.
From the symmetries (\ref{e7}), we know that $\det\Phi_+(\pm z_j)=0,~\det\Phi_-^{-1}(\pm\bar{z}_l)=0,~j,l=1,\cdots,N$.

For convenience, we introduce the notations
$$k_{2j}=z_j,\quad k_{2j-1}=-z_j.$$ 
Then the soliton matrix can be written in the form
\begin{equation}\label{e10}
\Gamma(n,z)=\chi_{2N}\chi_{2N-1}\cdots\chi_2\chi_1,
\end{equation}
where
\begin{equation}\label{e11}
\chi_l=I-\frac{k_l-\bar{k}_l}{z-\bar{k}_l}P_l, \quad \chi_j^{-l}=I+\frac{k_j-\bar{k}_j}{z+k_j}P_j,\quad
P_j=\frac{|j\rangle\langle j|}{\langle j|j\rangle},
\end{equation}
with the eigenvector $\langle j|=|j\rangle^\dag$ and $|j\rangle$ solves $\Phi_+(n,k_j)|j\rangle=0$.
In this case, one may find that $|2j\rangle=\sigma_3|2j-1\rangle$ and $P_{2j}=\sigma_3P_{2j-1}\sigma_3$.
Hence the product $\Phi_+(z)\chi_j^{-1}(z)$ is regular at the point $k_j$ and $\chi_l(z)\Phi_-^{-1}(z)$ is regular at $\bar{k}_l$.

Since $\Phi_+(n,z)$ solves the linear system (\ref{e1}) and (\ref{e2}), we know that the eigenvector $|j\rangle$
takes the form
\begin{equation}\label{e12}
|j\rangle(n,t)=E^n(k_j)e^{\omega_j\sigma_3t}|j_0\rangle,\quad \omega_j=\omega(k_j).
\end{equation}

The regular RH problem can be derived similarly as (\ref{c3a}) and (\ref{c3}), where the soliton matrix $\Gamma$ has the following decomposition
\begin{equation}\label{e13}
\Gamma=I-\sum\limits_{j,l=1}^{2N}\frac{1}{z-\bar{k}_l}|j\rangle(D^{-1})_{jl}\langle l|,\quad D_{lj}=\frac{\langle l|j\rangle}{k_j-\bar{k}_l}.
\end{equation}

For $N=1$, we take $z_1=e^{\xi+i\eta}$, that is $k_2=z_1,k_1=-z_1$,
\begin{equation}\label{e14}
|2\rangle=\left(\begin{array}{c}
e^{\theta+i\phi}\\
e^{-(\theta+i\phi)}
\end{array}\right),\quad |1\rangle=\sigma_3|2\rangle,
\end{equation}
where
\begin{equation}\label{e15}
\theta=n\xi+\sinh2\xi\cos2\eta t+\alpha,\quad
\phi=n\eta+\cosh2\xi\sin2\eta t+\beta.
\end{equation}
In this case, the soliton matrix takes the form
\begin{equation}\label{e16}
\Gamma(n,t,z)=I-\frac{D_-}{z-\bar{z}_1}-\frac{D_+}{z+\bar{z}_1},
\end{equation}
where
\begin{equation}\label{e17}
D_-=\frac{z_1-\bar{z}_1}{2}\left(
\begin{matrix}
\frac{e^{2\theta}}{z_1e^{-2\theta}+\bar{z}_1e^{2\theta}}&\frac{e^{2i\phi}}{z_1e^{-2\theta}+\bar{z}_1e^{2\theta}}\\
\frac{e^{-2i\phi}}{z_1e^{2\theta}+\bar{z}_1e^{-2\theta}}&\frac{e^{-2\theta}}{z_1e^{2\theta}+\bar{z}_1e^{-2\theta}}
\end{matrix}
\right), \quad D_+=-\sigma_3D_-\sigma_3.
\end{equation}

From (\ref{e16}) and (\ref{e17}), we have
\begin{equation}\label{e18}
\Gamma(n,z=0)=\frac{z_1}{\bar{z}_1}\left(
\begin{matrix}
\frac{z_1e^{2\theta}+\bar{z}_1e^{-2\theta}}{z_1e^{-2\theta}+\bar{z}_1e^{2\theta}}&0\\
0&\frac{z_1e^{-2\theta}+\bar{z}_1e^{2\theta}}{z_1e^{2\theta}+\bar{z}_1e^{-2\theta}}
\end{matrix}
\right),
\end{equation}
and
\begin{equation}\label{e19}
\Gamma^{(1)}(n)=-(z_1^2-\bar{z}_1^2)\left(
\begin{matrix}
0&\frac{e^{2i\phi}}{z_1e^{-2\theta}+\bar{z}_1e^{2\theta}}\\
\frac{e^{-2i\phi}}{z_1e^{2\theta}+\bar{z}_1e^{-2\theta}}&0
\end{matrix}
\right),
\end{equation}
where $\Gamma^{(1)}(n)$ is defined by the asymptotic behavior
\begin{equation}\label{e20}
\Gamma(n,z)=I+z^{-1}\Gamma^{(1)}(n)+O(z^{-2}),\quad z\rightarrow\infty.
\end{equation}

Next we will give the solution of dmKdV equation (\ref{a1}) for $N=1$.
To this end, we take the asymptotic behavior of the sectionally holomorphic $\Phi_+(n,z)$ as
$$\Phi_+(n,z)=\Phi_+^{(0)}(n)+z^{-1}\Phi_+^{(1)}(n)+O(z^{-2}),\quad z\rightarrow\infty,$$
which together with $\Phi_+(n,z)=\Phi_+^{(0)}(n)\Gamma(n,z)$ imply that $\Phi_+^{(1)}(n)=\Phi_+^{(0)}(n)\Gamma^{(1)}(n)$.
Note that the solution $u(n,t)$ can be rewritten as
\begin{equation}\label{e21}
u(n,t)=-\frac{\Phi_{+12}^{(1)}(n)}{\Phi_{+22}^{(0)}(n)}=-\frac{\nu_-(n)}{\nu_+(n)}\Gamma_{12}^{(1)}(n),
\end{equation}
in view of (\ref{e6}), (\ref{e8}) and the definition of $\Phi_+(n,z)$.
On the other hand, since $S^\dag(\bar{z})=S^{-1}(z), \det S(z)=1$,  and $a_+(z)=\det\Phi_+(n,z)\rightarrow\nu_+(n)\nu_-(n),~z\rightarrow\infty$,
as well as $G(z)=I$ in the RH problem as in (\ref{c1}), then
\begin{equation}\label{e22}
\nu=\nu_+(n)\nu_-(n)=1.
\end{equation}
Taking notice of the RH problem reduces to $\Phi_+(n,z)=\Phi_-(n,z)$, which allows us to discuss the
asymptotic behavior of $\Phi_+(n,z)$ near $z=0$,
\begin{equation}\label{e23}
\Phi_+(n,z)\rightarrow\tilde{\Phi}_-^{(0)}(n),\quad z\rightarrow 0,
\end{equation}
where $\tilde{\Phi}_-^{(0)}(n)$ defined in (\ref{e9}). Now using again $\Phi_+(n,z)=\Phi_+^{(0)}(n)\Gamma(n,z)$, we know that
\begin{equation}\label{e24}
\Gamma(n,z=0)=\left(\begin{matrix}
\nu_-^{-2}(n)&0\\
0&\nu_+^{-2}(n)
\end{matrix}\right).
\end{equation}

Hence the solution $u(n,t)$ can be reconstructed from (\ref{e24}), (\ref{e22}) and (\ref{e21})
\begin{equation}\label{e21}
u(n,t)=-\Gamma_{22}(n,z=0)\Gamma_{12}^{(1)}(n),
\end{equation}
For $N=1$, we have the one soliton solution of dmKdV equation (\ref{a1}) from (\ref{e18}) and (\ref{e19}) as
\begin{equation}\label{e21}
u(n,t)=e^{2\xi}\sinh2\xi{\rm sech}(2\theta+\xi),
\end{equation}
in terms of the assumption $\eta+2\phi=0$, where $\theta$ and $\phi$ are defined in (\ref{e15}).

\setcounter{equation}{0}
\section{Gauge transformation}
In this section, we discuss the Gauge transformation between the plus type dmKdV equation and the minus one.
Here we confine ourselves to the system (\ref{a4}).
In \cite{G-G}, the minus type dmKdV equation is the compatibility
condition of the Lax pair
\begin{equation}\label{f1}
\chi(n+1)=\tilde{L}_n\chi(n),\quad
\chi_t(n)=\tilde{M}_n\chi(n),
\end{equation}
where $\tilde{L}_n=(I+U_n)Z, \tilde{M}_n=k\sigma_3+\tilde{U}_n$ and
\begin{equation}\label{f2}
\begin{aligned}
&U_n=\left(\begin{matrix}
0&\tilde{u}(n)\\
\tilde{u}(n)&0
\end{matrix}\right),\\
&\tilde{U}_n=U_n+Z^{-1}U_{n-1}Z-U_nU_{n-1}.
\end{aligned}
\end{equation}

We let $\chi(n)=G_n\psi(n)$. If $\psi(n)$ and $Q_n$ solve the linear equations (\ref{a4}), then
\begin{equation}\label{f2a}
\begin{aligned}
\tilde{L}_n=G_{n+1}L_nG_n^{-1},\\
\tilde{M}_n=G_{n,t}G_n^{-1}+G_nM_nG_n^{-1},
\end{aligned}
\end{equation}
and
\begin{equation}\label{f3}
\tilde{u}(n)=-iu(n-2),
\end{equation}
where $L_n=(I+Q_n)Z, M_n=k\sigma_3+\tilde{Q}_n-Q_nQ_{n-1}$ and
\begin{equation}\label{f4}
G_n=\rho_n^\pm\left(\begin{matrix}
1-iu(n-1)\tilde{u}(n)\lambda&-u(n-1)-i\tilde{u}(n)\lambda\\
-\tilde{u}(n)\lambda+iu(n-1)\lambda^2&u(n-1)\tilde{u}(n)\lambda+i\lambda^2
\end{matrix}\right),
\end{equation}
with
$$\rho_n^+=\prod\limits_{k=n}^{+\infty}\frac{1+u^2(k)}{1-\tilde{u}^2(k)},\quad\rho_n^-=\prod\limits_{k=-\infty}^{n-1}\frac{1-\tilde{u}^2(k)}{1+u^2(k)}.$$
Indeed, $\tilde{L}_n=G_{n+1}L_nG_n^{-1}$ implies that the matrix $G_n$ can be represented in the following form
$$G_n=\left(\begin{matrix}
a_0(n)+a_1(n)\lambda&b_0(n)+b_1(n)\lambda\\
c_1(n)\lambda+c_2(n)\lambda^2&d_1(n)\lambda+d_2(n)\lambda^2
\end{matrix}\right).$$
Then we have a set of equations
\begin{subequations}
\begin{equation}\label{f5}
a_0(n+1)(1+u^2(n))=a_0(n)(1-\tilde{u}^2(n)),
\end{equation}
\begin{equation}\label{f6}
b_0(n+1)=-u(n)a_0(n+1),\quad c_1(n)=-\tilde{u}(n)a_0(n),
\end{equation}
\begin{equation}\label{f7}
\tilde{u}(n)b_0(n)=u(n)c_1(n+1)+d_1(n+1)-d_1(n),
\end{equation}
\begin{equation}\label{f8}
d_2(n+1)(1+u^2(n))=d_2(n)(1-\tilde{u}^2(n)),
\end{equation}
\begin{equation}\label{f9}
b_1(n)=-\tilde{u}(n)d_2(n), \quad c_2(n+1)=u(n)d_2(n+1),
\end{equation}
\begin{equation}\label{f10}
a_1(n+1)-a_1(n)-u(n)b_1(n+1)=\tilde{u}(n)c_2(n),
\end{equation}
\begin{equation}\label{f11}
b_0(n)=u(n)a_1(n+1)+b_1(n+1)-\tilde{u}(n)d_1(n),
\end{equation}
\begin{equation}\label{f12}
c_2(n)=-\tilde{u}(n)a_1(n)+c_1(n+1)-u(n)d_1(n+1).
\end{equation}
\end{subequations}
Hence, equation (\ref{f5}) implies $a_0(n)=\rho_n^\pm$, then $b_0, c_1$ and $d_1$ can be obtained from (\ref{f6}) and (\ref{f7}).
We take $d_2(n)=\alpha\rho_n^\pm$ by (\ref{f8}), then by (\ref{f9}) and (\ref{f10}), $b_1, c_2$ and $a_1$ is at hand,
where $\alpha$ is some constant. Thus $G_n$ in (\ref{f4}) is obtained.
In addition, the last two equation (\ref{f11}) and (\ref{f12}) product
\begin{equation}\label{f13}
\alpha\tilde{u}(n+1)=u(n-1),\quad \tilde{u}(n+1)=-\alpha u(n-1),
\end{equation}
in view of the identity $\rho_{n+1}^\pm(1+u^2(n))=\rho_n^\pm(1-\tilde{u}^2(n))$. Thus (\ref{f3}) is proven.
It is remarked that the second equation of (\ref{f2a}) is valid for $G_n$ given by (\ref{f4}), since
the gauge transformation between (\ref{a4}) and (\ref{f1}) implies
$$\tilde{L}_{n,t}-\tilde{M}_{n+1}\tilde{L}_n+\tilde{L}_n\tilde{M}_n=G_{n+1}(L_{n,t}-M_{n+1}L_n+L_nM_n)G_n^{-1}.$$
In this equation, $L_{n,t}-M_{n+1}L_n+L_nM_n=0$ implies the plus type equation of (\ref{a1}),
while $\tilde{L}_{n,t}-\tilde{M}_{n+1}\tilde{L}_n+\tilde{L}_n\tilde{M}_n=0$ gives the minus one.

We note that the gauge transformation about the similar problem of (\ref{a2}) gives rise to
\begin{equation}\label{f14}
\tilde{u}(n)=-iu(n-1).
\end{equation}
Hence the solutions of minus type dmKdV equation can be obtained by (\ref{f3}) or (\ref{f14}) from the given plus ones.
It is interesting to remark that the soliton solutions to minus type dmKdV equation can be also obtained without
needing to consider the nonvanishing boundary conditions \cite{SC}, and the solutions are complex value by equation (\ref{f3}) or (\ref{f14}) for above two cases.

\setcounter{equation}{0}
\section*{Acknowledgments}
Projects 11001250 and 11171312 are supported by the National Natural Science Foundation of China.
The work of JY Zhu is partially supported by the Foundation for Young Teachers in Colleges and Universities of Henan Province.

\end{document}